\documentclass[aps,twocolumn,pre, superscriptaddress]{revtex4-2}

\usepackage[english]{babel}
\usepackage{amsmath, amssymb, mathtools, physics}
\usepackage{braket}
\usepackage{ifthen}
\usepackage{graphicx,float}
\usepackage{multirow}
\usepackage[RGB,dvipsnames]{xcolor}
\definecolor{dodgerblue}{rgb}{0.12, 0.56, 1.0}
\definecolor{darkcyan32144140}{RGB}{32,144,140}
\definecolor{darkslateblue5294141}{RGB}{52,94,141}      
\definecolor{darkslateblue6467135}{RGB}{64,67,135}
\definecolor{darkslateblue7235116}{RGB}{72,35,116}      
\definecolor{darkslategray38}{RGB}{38,38,38}
\definecolor{greenyellow18922238}{RGB}{189,222,38}      
\definecolor{indigo68184}{RGB}{68,1,84}                 
\definecolor{lavender234234242}{RGB}{234,234,242}
\definecolor{mediumseagreen34167132}{RGB}{34,167,132}   
\definecolor{mediumseagreen68190112}{RGB}{68,190,112}
\definecolor{teal41120142}{RGB}{41,120,142}
\definecolor{yellowgreen12120981}{RGB}{121,209,81}
\colorlet{bluedarkslateblue5294141}{blue!50!darkslateblue5294141}
\newcommand{\re}[1]{{\leavevmode#1}}

\newcommand{\eprAB}{\hat{\sigma}_{A,B}}
\newcommand{\qualityAB}{\mathcal{Q}_{A,B}}

\newcommand{\turblock}{bara15,ging16,hwan18,dech18,bara19,koyu20,liu19,otsu20,vo20,song21,rold21,yosh21,vu23a}
\newcommand{\kldblock}{mart19,skin21a,vdm22,haru22,vdm22b,maie24,haru24b}
\newcommand{\corrblock}{dech21,shir23a,dech23,gu24,dieb25a,stut25,dech23a,vanv24,gu26,chen26,asly26}
\newcommand{\othersblock}{dite24,piet24,dite25}
\usepackage{hyperref}
\usepackage[nameinlink]{cleveref}
\usepackage{enumitem}
\usepackage{tikz}
\usepackage{bm}
\usetikzlibrary{external,cd}
\tikzsetexternalprefix{tikz/}
\usepackage{pgfplots}

\usetikzlibrary{automata, arrows.meta, positioning, calc,shapes,math,decorations.markings}

\begin{document}
%
\title{Lower bounds on entropy production from dynamical correlation functions}
\author{Paul Raux}
 \thanks{These authors contributed equally}
  \author{Alexander M. Maier}
 \thanks{These authors contributed equally}
\author{Udo Seifert}

\affiliation{%
 II. Institut für Theoretische Physik, Universität Stuttgart, 70550 Stuttgart, Germany
}%
\date{\today}

\begin{abstract}
Entropy production is a key property in stochastic thermodynamics. For partially observed and coarse-grained systems, its inference is challenging and typically rests on proven lower bounds. We derive two versions of such bounds based on the asymmetry of experimentally accessible two-time correlation functions of coarse-grained state observables. For non-equilibrium steady states, the bound is valid for arbitrary correlation lag. For time-dependent processes, it requires the limit of vanishing lag. These bounds hold true for any system that follows either a Markovian dynamics or a coupled set of overdamped Langevin equations on some underlying, unobservable level of description. We illustrate the bounds for both types of dynamics and discuss their optimization and potential tightness.

\end{abstract}
\maketitle

\section{Introduction}
Stochastic thermodynamics is a well-established framework describing small systems for which thermal noise plays a non-negligible role \cite{seki10,jarz11,peli21,shir23,seif25}.
In experiments involving such systems, limited resolution of measurement devices typically leads to incomplete observations \cite{espo12}. The growing field of thermodynamic inference \cite{seif19} tackles this situation  by providing non-invasive and model-free methods to infer thermodynamic properties, such as affinities \cite{piet16,ohga23,lian23}, or topologies \cite{maie25,zhao25}. In particular, the entropy production rate (EPR), which quantifies the irreversibility of a system, is central in non-equilibrium thermodynamics and methods to infer it have attracted particular attention \cite{seif25a,ghos26}. This inference is challenging since all out-of-equilibrium degrees of freedom contribute. If some of these degrees of freedom are hidden, lower bounds are typically the best one can ask for \cite{\turblock,\kldblock,\corrblock,\othersblock}.

Extant lower bounds on the EPR can roughly be classified into three categories. One of the most widely explored lower bounds on the EPR is the thermodynamic uncertainty relation (TUR) \cite{bara15,ging16,hwan18,dech18,bara19,koyu20,liu19,otsu20,vo20,song21,rold21,yosh21,vu23a}. This inequality relates the precision of a detectable fluctuating current, the squared mean divided by the variance of this current, to entropy production.
Another type of bound on the EPR is strictly based on path weights of observable coarse-grained trajectories. These estimators determine the asymmetry between the path weight for a trajectory and its time reverse in the form of a Kullback-Leibler divergence \cite{mart19,skin21a,vdm22,haru22,vdm22b,maie24,haru24b}.

Since correlation functions are often determined in experiments across statistical physics and other disciplines \cite{zwan65,elso11,grig21,ange22,nett24,mack24,fran26,cont01,box08,chak11,chak11a,cohe11,stet16,yuan24}, methods for thermodynamic inference making use of them seem to be highly attractive. Indeed, estimators based on correlations constitute a third group of thermodynamic bounds sparking interest in recent years \cite{dech18,dech21,dech23,gu24,dieb25a,stut25,shir23a,ohga23,dech23a,lian23,vanv24,gu26,chen26,asly26}. These results range from improved versions of the TUR \cite{dech18,dech21} to genuine thermodynamic correlation inequalities \cite{dech23,gu24,dieb25a,stut25} and inequalities relating to correlation times \cite{dech23a}. Motivated by these considerations, we follow this recent line of research aimed at inferring thermodynamic quantities from correlation functions \cite{ohga23,dech23,dech23a,lian23,gu24,vanv24,dieb25a,stut25,gu26,chen26}. \re{Some of these methods target cycle affinities \cite{ohga23,lian23}. Bounds on the EPR typically hold in the limit of vanishing lag \cite{shir23a,gu24}, rely on integrated correlation functions \citep{asly26}, or involve spectral properties of the Markov generator \cite{vanv24}.}

The recently reported lower bound on the EPR in Ref.~\cite{chen26} uses correlation functions between normalized state observables effectively as a coarse-grained version of path weights for systems in a non-equilibrium steady state (NESS). By contrast, we allow for correlation functions between arbitrary, possibly coarse-grained, state observables. We consider the correlation function between observables $A$ at time $t$ and $B$ at a later time $t+\tau$. Moreover, we will prove a lower bound for the EPR based on correlation functions for time-dependent processes like relaxation or periodic driving.

This paper is organized as follows. In Sec.~\ref{sec : set-up}, we describe the setup. We then derive lower bounds on the EPR for systems in a NESS and for time-dependent processes in Sec.~\ref{sec : lower bounds on the EPR}. We discuss the conditions for which the bounds are saturated in Sec.~\ref{sec : optimal bound}. Moreover, we propose a way to optimize the bound in a NESS. Finally, in Sec.~\ref{sec : illustration}, we illustrate our results with simple examples.

\section{Setup}
\label{sec : set-up}

We assume that the system under observation is Markovian on some microscopic level of description that can be represented by the random variable $X_t$. Without loss of generality, we focus on discrete-state  Markov networks. Since overdamped Langevin processes can be obtained as the continuous-state-space limit of a discrete variable, our results apply in this setting as well. For a system with $N$ states, we denote with $\bm p(t)= (p_1(t), \dots, p_N(t))^\intercal$ the probability distribution on this set. Its evolution is given by the master equation
\begin{equation}
\frac{d\bm p(t)}{dt} = \bm L(t) \bm p(t)
\label{eq : master equation}
\end{equation} 
where the Markov generator $\bm L(t)$ is, in general, time-dependent. We assume $\bm L(t)$ to be irreducible, i.e., every state to be reachable from any other state. Its elements read
\begin{equation}
L_{ji}(t) = k_{ij}(t) -\delta_{ij}\sum_l k_{il}(t),
\label{eq : definition of L}
\end{equation}
where the transition rate $k_{ij}(t)$ for transitions from state $i$ to state $j$ is positive if $k_{ji}(t)$ is positive and vice versa. This condition enforces local-detailed balance for all pairs of transitions, which reduces to detailed balance if the system is in equilibrium.

The formal solution of the master equation \eqref{eq : master equation} is
\begin{equation}
  \bm p(t_2)=\bm P(t_2-t_1,t_1)\bm p(t_1)
\end{equation} 
with the propagator
\begin{equation}
  \bm P(t_2-t_1,t_1) = \overrightarrow{\exp}\left[\int_{t_1}^{t_2}\bm L(s) ds\right],
\end{equation}
which is a time-ordered exponential $\overrightarrow{\exp}[.]$ in the general case.
Its element $P_{ji}(t_2-t_1,t_1)\equiv(\bm P(t_2-t_1,t_1))_{ji}$ is the conditional probability for the system to be in state $j$ at time $t_2$ given that it is in $i$ at $t_1$.

The instantaneous mean EPR \cite{peli21,shir23,seif25}
\begin{equation}
    \sigma(t) \equiv \frac{1}{2}\sum_{ij}j_{ij}(t)\ln\frac{p_i(t)k_{ij}(t)}{p_j(t)k_{ji}(t)}
    \label{eq : true EPR}
\end{equation}
can be expressed using products of the net current from state $i$ to state $j$,
\begin{equation}
        j_{ij}(t)\equiv p_i(t)k_{ij}(t)- p_j(t)k_{ji}(t),
\end{equation}
and its conjugated affinity.
We aim to find a lower bound on $\sigma(t)$ by relying only on experimentally accessible correlation functions between two state observables $A$ and $B$. In general, the observed trajectories may be coarse-grained in which case not all states are resolved individually. The resulting $\Omega\leq N$ lumped states will be called meso-states. For all states $i$ of the underlying network being part of a meso-state $\alpha$, we then have
\begin{equation}
A_i = A_\alpha \qquad\text{and}\qquad B_i = B_\alpha.
\end{equation}
From now on, we refer to meso-states using greek indices while we keep using roman indices for states of the underlying network.

We focus on the time asymmetry of the two-point correlation function
\begin{align}
    C_{A,B}(t,\tau)& \equiv\langle A(t)B(t+\tau) \rangle= \sum_{\alpha \beta} A_\alpha B_\beta  \nu_{\alpha\beta}(t,\tau)\label{eq : correlation function}
\end{align}
given by
\begin{align}
\chi_{A,B}(t,\tau)\equiv C_{A,B}(t,\tau) - C_{B,A}(t,\tau).
\label{eq : time asymmetry}
\end{align}
In definition \eqref{eq : correlation function}, we have introduced the joint probability
\begin{equation}
\nu_{\alpha\beta}(t,\tau)\equiv P(\alpha,t;\beta,t+\tau)
\end{equation}
to find the system in state $\alpha$ at time $t$ and in $\beta$ at $t+\tau$. It is the sum of probabilities $\nu_{ij}(t,\tau)\equiv P_{ji}(\tau,t)p_i(t)$ of all trajectories on the underlying level that lead to the same observed trajectory, i.e.,
\begin{equation}
  \nu_{\alpha\beta}(t,\tau) = \sum_{i\in \alpha, j\in \beta} \nu_{ij}(t,\tau). \label{eq : nu alpha beta}
\end{equation}
Defining the asymmetric \re{quantity}
\begin{align}
  d_{\alpha\beta}^{A,B} \equiv A_\alpha B_\beta - B_\alpha A_\beta,
  \label{eq : asymmetric increment}
\end{align}
\re{which will act like an increment of a current later on,} and using Eqs.~\eqref{eq : correlation function} and \eqref{eq : nu alpha beta}, we can express the correlation asymmetry \eqref{eq : time asymmetry} as
\begin{equation}
    \chi_{A,B}(t,\tau) = \sum_{\alpha\beta} d_{\alpha\beta}^{A,B}\nu_{\alpha\beta}(t,\tau). \label{eq:chi_nu}
\end{equation}

Before further rewriting the correlation asymmetry to prepare for deriving entropy estimators, we introduce the current-like quantities
\begin{align}
   \mathcal{J}_{ij}(t,\tau)&\equiv \nu_{ij}(t,\tau)-\nu_{ji}(t,\tau) \label{eq : Jij}
\end{align}
and
\begin{align}
  \mathcal{J}_{\alpha\beta}\re{(t,\tau)} &\equiv \sum_{i\in \alpha, j\in \beta}\mathcal{J}_{ij}(t,\tau), \label{eq : Jalphabeta}
\end{align}
and, in analogy to them, traffic-like quantities
\begin{align}
  \mathcal{T}_{ij}(t,\tau)&\equiv \nu_{ij}(t,\tau)+\nu_{ji}(t,\tau) \label{eq : definition of the traffic}
\end{align}
and
\begin{align}
  \mathcal{T}_{\alpha\beta}(t,\tau)&\equiv\sum_{i\in \alpha, j\in \beta}\mathcal{T}_{ij}(t,\tau)
\end{align}
for transitions between states of the underlying network and between the coarse-grained meso-states, respectively. Using the current-like quantities in \eqref{eq : Jij} and \eqref{eq : Jalphabeta} allows us to rewrite the correlation asymmetry \eqref{eq:chi_nu} in terms of products of quantities that are antisymmetric under the exchange of indices,
\begin{equation}
    \chi_{A,B}(t,\tau) = \frac{1}{2}\sum_{\alpha\beta} d_{\alpha\beta}^{A,B}\mathcal{J}_{\alpha\beta}(t,\tau).
    \label{eq : asymmetry prop current-like}
\end{equation}

\section{Lower bounds for the EPR}
\label{sec : lower bounds on the EPR}
We now turn to the derivation of lower bounds on the EPR. The first subsection is dedicated to the derivation of an intermediate bound based on $\chi_{A,B}(t,\tau)$ valid for any, possibly time-dependent, process. In the second subsection, we specialize the latter inequality in a NESS to obtain a lower bound on the EPR there. Finally, in the third subsection, we specialize the intermediate inequality to systems with time-dependent driving and to relaxation processes to obtain a lower bound on their instantaneous EPR.

\subsection{Upper bound on the correlation asymmetry}
\label{subsec : upper bound on the two time correlation asymmetry}
First, we multiply each term in the correlation asymmetry \eqref{eq : asymmetry prop current-like} by one by introducing the factor $\sqrt{\mathcal{T}_{\alpha\beta}/\mathcal{T}_{\alpha\beta}}$. Second, we apply the Cauchy-Schwarz inequality, $(\sum_l a_lb_l)^2\leq \sum_la_l^2\sum_mb_m^2$, to this modified version of Eq. \eqref{eq : asymmetry prop current-like}, which leads to
\begin{align}
  4\left[ \chi_{\re{A,B}}(t,\tau)\right]^2 &\leq 2\mathcal{D}_{A,B}(t,\tau) \mathcal{F}(t,\tau) \label{eq : Cauchy-Schwarz step}
\end{align}
with
\begin{align}
  \mathcal{D}_{A,B}(t,\tau) &\equiv \frac{1}{2}\sum_{\alpha\beta} \left[{d_{\alpha\beta}^{A,B}}\right]^2\mathcal{T}_{\alpha\beta}(t,\tau) \label{eq:Fone}
\end{align}
and
\begin{align}
  \mathcal{F}(t,\tau) &\equiv \sum_{\alpha\beta}   \mathcal{J}_{\alpha\beta}(t,\tau)^2/\mathcal{T}_{\alpha\beta}(t,\tau). \label{eq:Ftwo}
\end{align}
Since we can express the sum \eqref{eq:Fone} in terms of correlation functions,
\begin{align}
      \mathcal{D}_{A,B}(t,\tau) =\; &C_{A^2,B^2}(t,\tau) + C_{B^2,A^2}(t,\tau) \notag \\ &- 2 C_{\re{AB,AB}}(t,\tau),
      \label{eq : denominator expression}
\end{align}
it is operationally accessible.

The sum \eqref{eq:Ftwo} is upper bounded by a coarse-grained EPR-like intermediate
\begin{align}
 \mathcal{F}(t,\tau) \leq \frac{1}{2}\sum_{\alpha\beta} \mathcal{J}_{\alpha\beta}(t,\tau)\ln\frac{ \nu_{\alpha\beta}(t,\tau) }{ \nu_{\beta\alpha}(t,\tau) }\equiv\hat{\Sigma}^\text{cg}(t,\tau)
 \label{eq : EPR like estimator}
\end{align}
by virtue of the inequality \re{$(a-b)^2/(a+b)\leq (a-b)\ln(a/b)/2$, which results from Jensen's inequality for arbitrary $a,b\geq 0$.} This EPR-like intermediate is the Kullback-Leibler divergence between the flow-like probabilities $\nu_{\alpha\beta}(t,\tau)$ and the corresponding ones for the reversed trajectory $\beta\to\alpha$. Since these $\nu_{\alpha\beta}(t,\tau)$ are a sum of joint probabilities between states of the underlying Markov network, we use the log-sum inequality, $\sum_i a_i\ln(a_i/b_i) \geq \sum_i a_i\ln(\sum_ja_j/\sum_lb_l)$ for positive numbers $a_i$ and $b_i$, to arrive at
\begin{equation}
  \hat{\Sigma}^\text{cg}(t,\tau)\leq \sum_{ij} \nu_{ij}(t,\tau)\ln\frac{ \nu_{ij}(t,\tau) }{ \nu_{ji}(t,\tau) } \equiv \hat{\Sigma}(t,\tau).
  \label{eq : EPR like estimator DKL not coarse grained}
\end{equation}

Combining inequality \eqref{eq : Cauchy-Schwarz step} with relations \eqref{eq : denominator expression} to \eqref{eq : EPR like estimator DKL not coarse grained} yields the inequality
\begin{equation}
    \hat \sigma_{A,B}(t,\tau)\equiv \frac{2}{\tau} \frac{\chi_{A,B}(t,\tau)^2}{\mathcal{D}_{A,B}(t,\tau)} \leq \frac{\hat{\Sigma}(t,\tau)}{\tau}.
    \label{eq : intermediate result}
\end{equation}
While this result is not yet a lower bound on the EPR in general, we can specialize it to one.

\subsection{NESS}\label{subsec : lower bound for the EPR}

We now assume that the system is in a NESS where the distribution $\bm p$ is time-independent. More generally, all quantities are independent of an initial time $t$ and depend, like the propagator $\bm P(\tau)$, only on the lag between two measurements.
Moreover, the intermediate $\hat{\Sigma}(t,\tau)$ defined in \eqref{eq : EPR like estimator DKL not coarse grained} is a coarse-grained mean entropy change
\begin{equation}
    \hat{\Sigma}(\tau)\equiv \sum_{ij} \nu_{ij}(\tau)\ln\frac{\nu_{ij}(\tau)}{\nu_{ji}(\tau)}.
 \label{eq : EPR like estimator NESS}
\end{equation}

The $\nu_{ij}(\tau)$ are given by a sum over path weights $\mathbb{P}[\gamma_{i\to j}(\tau)]$ of microscopic trajectories $\gamma_{i\to j}(\tau)$ running from state $i$ to state $j$ of fixed length $\tau$ as
\begin{align}
  \nu_{ij}(\tau) &=\sum_{\gamma_{i\to j}(\tau)}\mathbb{P}\left[\gamma_{i\to j}(\tau)\right] \label{eq : forward two points coarse graining}
\end{align}
and
\begin{align}
  \nu_{ji}(\tau) &=\sum_{\gamma_{i\to j}(\tau)}\tilde{\mathbb{P}}\left[\widetilde{\gamma_{i\to j}}(\tau)\right] = \sum_{\gamma_{j\to i}(\tau)}\mathbb{P}\left[\gamma_{j\to i}(\tau)\right]. \label{eq : backward two points coarse graining}
\end{align}
Here, we indicate time reversal with a tilde, which leaves $\mathbb{P}$ unaltered in the NESS. Expressing the $\nu_{ij}(\tau)$ in \eqref{eq : EPR like estimator NESS} via Eqs. \eqref{eq : forward two points coarse graining} and \eqref{eq : backward two points coarse graining} and applying the log-sum inequality once again yields\begin{equation}
    \hat{\Sigma}(\tau)\leq D_\text{KL}(\mathbb{P}\left[\gamma(\tau)\right]\vert\re{\vert} \tilde{\mathbb{P}}\left[\tilde\gamma(\tau)\right])=\tau \sigma.
    \label{eq : log sum inequality}
\end{equation}
Here, we use that the Kullback-Leibler divergence between the path weight of a microscopic trajectory and its time reverse in an interval of length $\tau$ is the average entropy production over this interval by virtue of a fluctuation theorem and that the EPR is constant in the NESS \cite{seif25}. Combined with inequality \eqref{eq : intermediate result}, this yields our first main result
\begin{equation}
    \hat\sigma_{A,B}(\tau) \equiv \frac{2}{\tau}\frac{\chi_{A,B}(\tau)^2}{\mathcal{D}_{A,B}(\tau)}\leq \sigma,
    \label{eq : main result NESS}
\end{equation}
which is a lower bound on the mean EPR $\sigma$ valid for any lag $\tau \geq 0$ in a NESS.

This result also applies to Markov processes in discrete time, e.g., resulting from stroboscopic observations. There, the number $n$ of time steps replaces $\tau$, i.e., state evolution is given by the propagator for one time step raised to the power $n$ instead of given by a time-dependent propagator, and $\sigma$ is the mean entropy change per time step. For these processes, the inequality corresponding to the bound \eqref{eq : log sum inequality} of the above derivation is saturated and we arrive at the bound \eqref{eq : main result NESS} where $\tau$ is replaced by $n$ and the interpretation of $\sigma$ is changed accordingly.

\subsection{Time-dependent processes}
For relaxation and time-dependently driven processes, time-reversed trajectories are inaccessible by observations of the forward dynamics alone. Hence, in this case, we cannot follow the previous derivation for a NESS. However, focussing on the limit of vanishing lag $\tau \rightarrow 0$, we do not need access to these trajectories of the time-reversed dynamics.

For small lag \re{and $i\neq j$}, the propagator $\boldsymbol{P}(\re{\tau,t})$ admits the expansion
\begin{equation}
  P_{ji}(\re{\tau,t}) =\delta_{ji} + k_{ij}(t)\tau +\mathcal{O}(\tau^2). \label{eq : small lag expansion of the propagator}
\end{equation}
This implies that $\nu_{ij}(t,\tau)$ can be expanded as
\begin{equation}
  \nu_{ij}(t,\tau)=p_i(t)\left[\delta_{ij} + \tau k_{ij}(t)+ O(\tau^2)\right]. \label{eq : small lag expansion of the micro proba}
\end{equation}
The vanishing-lag limit of $\hat{\Sigma}(t,\tau)$, based on these expansions, then reads
\begin{equation}
     \lim_{\tau \to 0}\frac{ \hat{\Sigma}(t,\tau)}{\tau}=\sum_{i j} p_i(t)k_{ij}(t)\ln\frac{p_i(t)k_{ij}(t)}{p_j(t)k_{ji}(t)} = \sigma(t).
    \label{eq : small lag exp EPR like estimator time dep}
\end{equation}
The equality to the time-dependent EPR \eqref{eq : true EPR} becomes apparent by considering index permutations and utilizing the asymmetry of the log ratios. Combining this last equality \eqref{eq : small lag exp EPR like estimator time dep} with inequality \eqref{eq : intermediate result} leads to our second main result
\begin{equation}
   \eprAB(t)\equiv    \lim_{\tau \to 0}  \hat \sigma_{A,B}(t,\tau) =\lim_{\tau\to 0} \frac{2}{\tau}\frac{\chi_{A,B}(t,\tau)^2}{\mathcal{D}_{A,B}(t,\tau)}\leq \sigma(t), \label{restdep:reslimit}
\end{equation}
which is a lower bound on the instantaneous mean EPR $\sigma(t)$ in general time-dependent processes.

The estimator $\eprAB(t)$ simplifies upon inspection of the terms in the numerator and the denominator. The asymmetry \eqref{eq : time asymmetry} in the numerator is proportional to $\mathcal{J}_{ij}(t,\tau)$. Their expansions for a small lag are
\begin{align}
  \mathcal{J}_{ij}(t,\tau) &= \tau j_{ij}(t) + \mathcal{O}(\tau^2)
\end{align}
and
\begin{align}
  \chi_{A,B} (t,\tau) &= \tau \partial_\tau  \chi_{A,B} (t,\tau)\vert_{\tau=0}+\mathcal{O}(\tau^2)
\end{align}
meaning that both vanish for $\tau=0$.
For small $\tau$, the denominator in \eqref{restdep:reslimit} reads
\begin{align}
  \mathcal{D}_{A,B}(t,\tau)=\tau  \partial_\tau [&C_{B^2,A^2}(t,\tau) + C_{A^2,B^2}(t,\tau)\notag\\
  &-2C_{BA,BA}(t,\tau) ]_{\tau=0} + \mathcal{O}(\tau^2).
\end{align}
Using these expansions, we can thus express the estimator \eqref{restdep:reslimit} in terms of the slope of correlation functions at vanishing lag,
\begin{align}
     &\eprAB(t)\notag\\
     &= \!\left.\frac{2\left[\partial_\tau  \chi_{A,B} (t,\tau)\right]^2}{\partial_\tau \left[C_{B^2,A^2}(t,\tau) + C_{A^2,B^2}(t,\tau)-2C_{BA,BA}(t,\tau) \right]}\right\vert_{\tau=0}\!. \label{restdep:resderivatives}
\end{align}
In the case of a NESS, this result is the $\tau\to 0$ limit of the bound \eqref{eq : main result NESS}.

\section{Tightness and optimization}
\label{sec : optimal bound}
It is natural to ask under which conditions the lower bounds on the EPR, Eqs.~\eqref{eq : main result NESS} and \eqref{restdep:reslimit}, can be saturated or, at least, become optimal. We start by examining the conditions in which the lower bounds recover the full EPR. Then, we show that the NESS estimator can be optimized analytically by shifting the state observables by constant values. For this discussion, we introduce the vectors $\bm A \equiv (A_1,\cdots A_\Omega)$ and $\bm B\equiv (B_1, \cdots B_\Omega)$, which collect the values taken by state observables on the $\Omega$ meso-states, as well as the increment matrix $\bm d^{A,B}$ with elements defined in Eq.\eqref{eq : asymmetric increment}.

\subsection{Tightness} \label{sec : saturation}
When deriving the bounds \eqref{eq : main result NESS} and \eqref{restdep:reslimit}, we have used a sequence of inequalities, i.e., the Cauchy-Schwarz inequality, the log inequality \re{$(a-b)^2/(a+b)\leq (a-b)\ln(a/b)/2$}, and the log-sum inequality.
The Cauchy-Schwarz inequality is saturated only if
\begin{equation}
    \mathcal{R}_{\alpha\beta}(t,\tau)\equiv\frac{\mathcal{J}_{\alpha\beta}(t,\tau)}{\mathcal{T}_{\alpha\beta}(t,\tau)} \propto d_{\alpha\beta}^{A,B}, 
    \label{eq : Cauchy Schwarz saturation condition}
\end{equation}
which requires proportionality between $d^{A,B}_{\alpha\beta}$ and the antisymmetric matrix element $\mathcal{R}_{\alpha\beta}(t,\tau)$. 
The increment matrix $\bm d^{A,B}$ is of rank two. Indeed, for non-zero state observables $A$ and $B$, the rank of $\bm d^{A,B}$ is constrained as  $\mathrm{rank}(\bm d^{A,B})= \mathrm{rank}(\bm A \bm B^\intercal - \bm B \bm A^\intercal)\leq \mathrm{rank}(\bm A \bm B^\intercal )+ \mathrm{rank}(\bm B \bm A^\intercal) $. Since $\bm A \bm B^\intercal$ and $\bm B \bm A^\intercal$ are outer products of non-zero vectors, they have rank $1$. Therefore $\mathrm{rank}(\bm d^{A,B})\leq 2$. Moreover, since $\bm d^{A,B}$ vanishes for co-linear vectors $\bm A$ and $\bm B$, any non-trivial increment has rank $2$. Since an antisymmetric matrix necessarily has an even rank \cite{axle15} and by excluding the trivial rank-zero case, in which $\mathcal{R}_{\alpha\beta}(t,\tau)=0$, we find that the saturation condition \eqref{eq : Cauchy Schwarz saturation condition} requires that $\mathcal{R}_{\alpha\beta}(t,\tau)$ be of rank $2$ and that it can be decomposed proportionally to $d^{A,B}_{\alpha\beta}$.

The log inequality \re{$(a-b)^2/(a+b)\leq (a-b)\ln(a/b)/2$} is saturated only for $a=b$. Thus, saturation occurs if all current-like \re{quantities} $\mathcal{J}_{ij}(t,\tau)$ \re{vanish,} which happens only in equilibrium. The log-sum inequality used in Eq.~\eqref{eq : EPR like estimator DKL not coarse grained} to account for the state coarse-graining is saturated in the special case in which there is no coarse-graining. This inequality was used a second time to conclude the proof in the NESS. For vanishing lag, the $\hat{\Sigma}(\tau)$ defined in inequality \eqref{eq : EPR like estimator DKL not coarse grained} reduces as in \eqref{eq : small lag exp EPR like estimator time dep} to the EPR. Therefore, inequality \eqref{eq : log sum inequality} is saturated there.

\subsection{Optimization in the NESS}\label{sec : optimization}
\subsubsection{Constant shifts}\label{sec : constant shifts}

We define the shifted observables
\begin{equation}
\bm A^\prime \equiv \bm A - s_A \mathbf{1}
\end{equation}
and 
\begin{equation}
\bm B^\prime \equiv \bm B - s_B \mathbf{1},
\end{equation}
where $s_A$ and $s_B$ are real. Since the value of the bound $\hat \sigma_{A,B}(\tau)$ changes under this transformation, we can optimize the estimator \eqref{eq : main result NESS} as a function of these shifts. Indeed, we now show that the correlation asymmetry \eqref{eq : asymmetry prop current-like} is unchanged under the transformation whereas the denominator $\mathcal{D}_{A^\prime,B^\prime}(\tau)$ admits a global minimum. 

The increment $d_{\alpha\beta}^{A,B}$ transforms as 
\begin{equation}
d_{\alpha\beta}^{A^{\prime},B^{\prime}} = d_{\alpha\beta}^{A,B} + s_A\Delta B_{\alpha\beta} - s_B\Delta A_{\alpha\beta}
\end{equation}
where $\Delta A_{\alpha\beta}=A_\alpha - A_\beta$ and $\Delta B_{\alpha\beta}=B_\alpha - B_\beta$. Plugging this expression into the asymmetry \eqref{eq : asymmetry prop current-like}, we obtain 
\begin{align}
\chi_{A^\prime,B^\prime}(\tau)=&\chi_{A,B}(\tau) \notag\\
&+  \re{\frac{1}{2}}\sum_{\alpha,\beta}\left(s_A\Delta B_{\alpha\beta}-s_B\Delta A_{\alpha\beta}\right)\re{\mathcal{J}}_{\alpha\beta}(\tau).
\end{align}
Using the antisymmetry of $J_{\alpha\beta}(\tau)$ under the exchange of its indices together with Kirchhoff's current law, $\sum_j J_{ij}(\tau)=0$, we find that the correlation asymmetry is invariant under this shift transformation.

However, the denominator \eqref{eq : denominator expression} depends on the square of $d_{\alpha\beta}^{A,B}$, which admits the quadratic form
\begin{equation}
\left(d_{\alpha\beta}^{A^{\prime},B^{\prime}} \right)^2=\left(d_{\alpha\beta}^{A,B} \right)^2 +\bm s^\intercal \bm v_{\alpha\beta} +\bm s^\intercal\bm h_{\alpha\beta}\bm s
\end{equation}
with
\begin{equation}
\bm v_{\alpha\beta} = 2d_{\alpha\beta}^{A,B}
\begin{pmatrix}
\Delta B_{\alpha\beta}\\
-\Delta A_{\alpha\beta}
\end{pmatrix},
\end{equation}
\begin{equation}
\bm h_{\alpha\beta} = \begin{pmatrix}
\Delta B_{\alpha\beta}^2 & \re{- \Delta B_{\alpha\beta}\Delta A_{\alpha\beta}} \\
 \re{- \Delta B_{\alpha\beta}\Delta A_{\alpha\beta}}& \Delta A_{\alpha\beta}^2
\end{pmatrix} 
\end{equation}
and $\bm s=(s_A,s_B)^\intercal$. Inserting this quadratic form into \eqref{eq : denominator expression} yields
\begin{align}
\mathcal{D}_{A^\prime,B^\prime}(\tau)
=
\mathcal{D}_{A,B}(\tau)
+
\frac{1}{2}\bm s^{\re{\intercal}}\bm{\mathcal{H}}(\tau)\bm s
+
\frac{1}{2}\bm s^{\re{\intercal}}\bm V(\tau).
\label{eq : shifted denominator}
\end{align}
Both the vector $\bm V(\tau)$ and the Hessian matrix $\bm{\mathcal{H}}(\tau)$ can be expressed solely in terms of one- and two-time correlation functions of the observables $A$ and $B$. The entries of the Hessian matrix,
\begin{align}
\bm{\mathcal{H}}(\tau)=
\begin{pmatrix}
\mathcal{H}_1(\tau) & \mathcal{H}_2(\tau)\\
\mathcal{H}_2(\tau) & \mathcal{H}_4(\tau)
\end{pmatrix},
\end{align}
are
\begin{align}
\mathcal{H}_1 (\tau)&= 4\left[\langle \bm B^2\rangle - C_{B,B}(\tau)\right],\label{eq : H1}\\
\mathcal{H}_2 (\tau)&= \re{2}\left[C_{A,B}(\tau)+C_{B,A}(\tau)-2\langle \bm A\bm B\rangle\right],\\
\mathcal{H}_4 (\tau)&= 4\left[\langle \bm A^2\rangle - C_{A,A}(\tau)\right].
\end{align}
The vector $\bm V\equiv (V_1,V_2)^\intercal$ has components
\begin{align}
V_1 (\tau)= 4 &\left[ C_{AB,B}(\tau)+C_{B,AB}(\tau)\right.\notag\\
&\left. -C_{B^2,A}(\tau)-C_{A,B^2}(\tau)\right]
\end{align}
and 
\begin{align}
V_2 (\tau)=4&\left[ C_{AB,A}(\tau)+C_{A,AB}(\tau) \right.\notag\\
&\left.- C_{A^2,B}(\tau)-C_{B,A^2}(\tau)\right].\label{eq : V2}
\end{align}
Furthermore, $\bm{\mathcal{H}}(\tau)$ and $\bm V(\tau)$ vanish for $\tau=0$. Since $\mathcal{D}_{A,B}(\tau)$ is non-negative by construction and a quadratic function of the shift vector $\bm s$, its Hessian matrix is positive definite for $\tau>0$. Consequently, $\mathcal{D}_{A^\prime,B^\prime}(\tau)$ admits a unique global minimum with respect to $\bm s$, implying that the estimator \eqref{eq : main result NESS} admits a unique global maximum under constant shifts. This maximum can be found analytically by minimizing the quadratic form \eqref{eq : shifted denominator}, which leads to the optimal shift 
\begin{equation}
\bm s^*\re{(\tau)} = - \re{\bm{\mathcal{H}}}^{-1}(\tau)\bm V(\tau)/2
\end{equation}
and to the optimized version of the lower bound \eqref{eq : main result NESS},
\begin{equation}
\re{\hat{\sigma}}_{A,B}^*(\tau)=\frac{2}{\tau}\frac{\chi_{A,B}(\tau)^2}{\mathcal{D}_{A,B}(\tau)-\re{\bm V^{\re{\intercal}}(\tau)\re{\bm{\mathcal{H}}}^{-1}(\tau)\bm V(\tau)/8}}. \label{eq : optimal estimator}
\end{equation}

\subsubsection{Finite lag}\label{sec : finite lag}

The estimator can also be optimised as a function of $\tau$. Since we have seen in Sec.~\ref{sec : saturation} that being close to equilibrium is part of the saturation conditions of the inequalities leading to the lower bound \eqref{eq : main result NESS} we expect the estimator there to be best in the limit $\tau\rightarrow 0$. Far from equilibrium, optimizing the estimator as a function of $\tau$ may lead to a maximum at finite $\tau$ as shown below. Whereas this optimization is in general analytically intractable, scanning over $\tau$ would yield the optimum with respect to $\tau$ in numerical or real experiments.
\begin{figure*}
    \centering
    \includegraphics[scale=1]{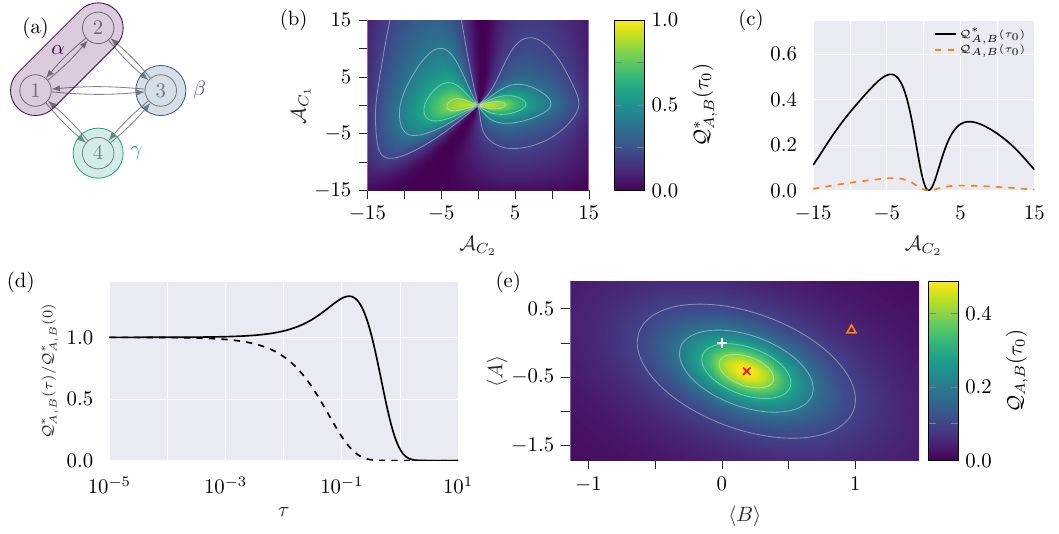}
     \caption{\re{Inferring entropy production in the NESS for a coarse-grained Markov network.
(a) Four-state network in which states 1 and 2 cannot be distinguished and observables are as given in Eq.~\eqref{eq:NESSobservables}.
(b) Quality factor $\mathcal Q^*_{A,B}(\tau_0)$ defined in \eqref{eq : quality factor optimal NESS} as a function of the cycle affinities $\mathcal{A}_{C_1}$ and $\mathcal{A}_{C_2}$ for fixed lag $\tau_0=10^{-3}$.
(c) Cut of the heatmap (b) for  $\mathcal A_{C_1}=6$ (solid) and the quality factor for the unshifted observables as a reference (dashed). 
(d) Ratio $\mathcal Q^*_{A,B}(\tau)/\mathcal Q^*_{A,B}(0)$ for $(\mathcal A_{C_1},\mathcal A_{C_2})=(6,-6)$ with $\mathcal Q^*_{A,B}(0)=0.49$ (dashed) and for $(\mathcal A_{C_1},\mathcal A_{C_2})=(-1.5,-0.38)$ with $\mathcal Q^*_{A,B}(0)=0.02$ (solid).
(e) Quality factor $\mathcal Q_{A,B}(\tau_0)$ defined in \eqref{eq : quality factor NESS} at $\mathcal{A}_{C_1}=6$ and $\mathcal{A}_{C_2}=-6$ versus the means of the shifted observables with optimum ($\times$) given by Eq.~\eqref{eq : optimal estimator}, with centred observables (+), and with unshifted observables ($\triangle$). 
For all panels, transition rates are given by equation \eqref{eq:NESSrates} and $(k_{12},k_{21},k_{13},k_{31},k_{14},k_{41})=(2,1,2,1,2,1)$.
}}
    \label{fig : figure diamond NESS}
\end{figure*}
\section{Illustrative examples}
\label{sec : illustration}
We now provide two examples to illustrate the performance of the bounds \eqref{eq : main result NESS} and \eqref{restdep:resderivatives}. First, we consider a four-state network with five pairs of transitions and two cycles. Second, we illustrate the lower bound \eqref{restdep:resderivatives} in a NESS for an overdamped Langevin particle on a ring.
For these illustrations, we define the quality factors
\begin{align}
  \qualityAB(\tau) &\equiv \eprAB(\tau)/\sigma, \label{eq : quality factor NESS} \\
  \re{\qualityAB^*(\tau)} &\re{\equiv \eprAB^*(\tau)/\sigma,} \label{eq : quality factor optimal NESS}\\
  \qualityAB(t) &\equiv \eprAB(t)/\sigma(t), \label{eq : quality factor tdep}
\end{align}
for a NESS\re{, with arbitrary and optimal shifts,} and for time-dependent processes, respectively, and
\begin{equation}
  \qualityAB \equiv \lim_{\tau\to 0} \qualityAB(\tau)
  \label{eq : quality factor tdep NESS}
\end{equation}
for a NESS.

\subsection{Discrete network}
\subsubsection{Non-equilibrium steady state}
Consider stationary driving of the Markov network in Fig. \ref{fig : figure diamond NESS}(a) with cycles $C_1 = 1\to 2\to 3\to 1$ and $C_2 = 1\to 4\to 3\to 1$. We vary the cycle affinities $\mathcal{A}_{C_1}$ and $\mathcal{A}_{C_2}$ using the rates
\begin{align}
k_{23} &= \exp(\mathcal{A}_{C_1}/2), \quad k_{32} = \exp(-\mathcal{A}_{C_1}/2),\notag\\
k_{43} &= \exp(\mathcal{A}_{C_2}/2), \quad  k_{34} = \exp(-\mathcal{A}_{C_2}/2). \label{eq:NESSrates}
\end{align}
The remaining rates are constants independent of $\mathcal{A}_{C_1}$ and $\mathcal{A}_{C_2}$. Furthermore, we assume that states $1$ and $2$ are coarse-grained into a single state $\alpha$ such that they cannot be distinguished based on the observed state variables chosen as
\begin{align}
\bm{A} = ( 1, 1, 0)^\intercal \qq{and} \bm{B} = ( 1, 0, 1)^\intercal. \label{eq:NESSobservables}
\end{align}

\re{Even with the two unresolved states in this example, the optimised lower bound \eqref{eq : optimal estimator} captures up to $93\,\%$ of the full EPR for sufficiently small affinities, as illustrated in Fig.~\ref{fig : figure diamond NESS} (b). Panel (c) shows a cut of the heat map (b) for a fixed value of $\mathcal{A}_{C_1}$ for which the optimal estimator reaches $\simeq 0.51$ since, for all values of $\mathcal{A}_{C_2}$, the system remains far from equilibrium and thus far from the saturation conditions discussed in Sec.~\ref{sec : saturation}. For reference, we present on the same graph the quality factor of the estimator based on the unshifted observables which reaches  $\simeq 0.05$. Moreover, the monotonicity of the quality factor as a function of $\tau$ depends on the cycle affinities as shown in panel (d). In this example, we observe the two cases described in Sec.~\ref{sec : finite lag}. We remark that the normalisation hides the fact that the quality factor of the non-monotonic case is very small, with $\simeq 0.03$ at the peak.}

\re{Additionally, Fig.~\ref{fig : figure diamond NESS} (e) illustrates that, for a given lag, the largest quality factor is found for observables with nonzero mean where it is $\simeq 0.49$. For centered observables, the quality factor is $\simeq 0.28$. For unshifted observables, it becomes $\simeq 0.05$. Thus, in the general case, setting the mean of observables to zero as can be done for other entropy estimators, e.g., for the one in Ref.~\cite{chen26}, can be detrimental for the bound \eqref{eq : main result NESS}.}

\subsubsection{Time-dependent driving}

\begin{figure}
    \centering
    \includegraphics[scale=1]{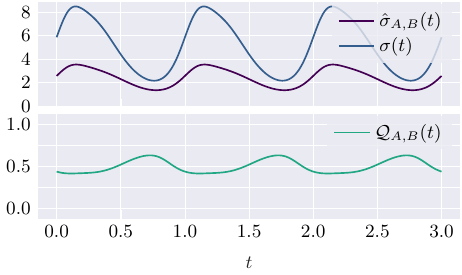}
   \caption{Inferring entropy production for the periodically driven, \re{coarse-grained} four-state Markov network from Fig. \ref{fig : figure diamond NESS} with observables \re{$(A_1,A_2,A_3,A_4) = (0,0,1,-2)$ and $(B_1,B_2,B_3,B_4) = (-1,-1,0,1)$, i.e., with $(A_\alpha,A_\beta,A_\gamma) = (0,1,-2)$ and $(B_\alpha,B_\beta,B_\gamma) = (-1,0,1)$}. \emph{Upper panel}: Time-dependent entropy production rate and its estimator \eqref{restdep:resderivatives}. \emph{Lower panel}: Quality factor \eqref{eq : quality factor tdep}. The transition rates are $\re{k_{34}}(t) = \exp\left[2+\sin(2\pi t)\right]$, $\re{k_{43}}(t) = \exp\left[-2 -\sin(2\pi t)\right]$, and \re{$(k_{12},k_{13},k_{14},k_{21},k_{23},k_{31},k_{32},k_{41}) = (1.1,1.5,0.9,1.9,2.3,1.0,0.7,4.1)$}.}
    \label{fig : figure diamond PSS}
\end{figure}

We now assume the four-state Markov network in Fig. \ref{fig : figure diamond NESS}(a) to be \re{coarse-grained} and driven periodically in time. The system thus reaches a periodic stationary state in the long-time limit, for which we use the bound \eqref{restdep:resderivatives} to get a time-dependent lower bound on the instantaneous EPR as shown in Fig. \ref{fig : figure diamond PSS}. \re{Even though the dynamics is partly hidden, we are able to infer roughly half of $\sigma(t)$ even using these time-independent observables.} Using suboptimal observables that are far from satisfying the condition \eqref{eq : Cauchy Schwarz saturation condition} at any point in time leads to a somewhat loose bound as this example demonstrates. To get a tighter estimate in the general time-dependent case with coarse-grained dynamics, we would need to use time-dependent observables. Optimizing the estimator for all times $t$ separately would then yield the optimized bound in the general time-dependent case.

\subsection{Overdamped Langevin dynamics}
\begin{figure*}
    \centering
    \includegraphics[scale=1]{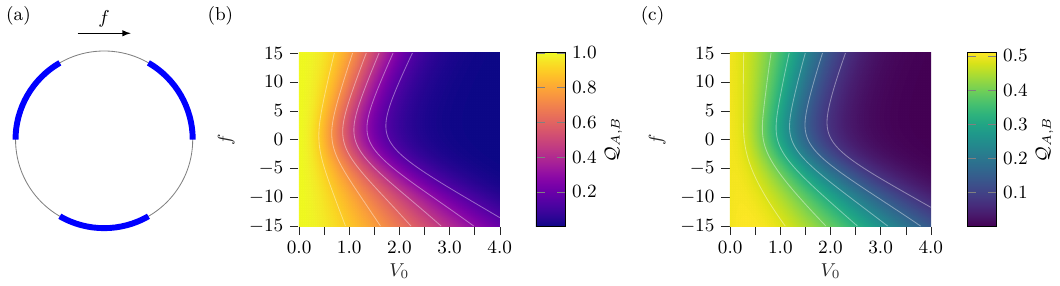}
    \caption{Estimating the entropy production for an overdamped Langevin particle on a one-dimensional ring with unit length subject to the potential \eqref{odeq:potential} and a non-conservative force $f$ for observables as given in \eqref{odeq:observables}.
      (a) Sketch of a one-dimensional ring. The intervalls $[1/12,1/4)$, $[5/12,7/12)$ and $[3/4,11/12)$ are highlighted in blue.
      (b) Quality factor $\qualityAB$ defined in \eqref{eq : quality factor tdep NESS} as a function of potential amplitude $V_0$ and driving force $f$ for the fully observable ring in (a).
      (c) Quality factor $\qualityAB$ as in (b) for the case in which only the blue regions of the ring in (a) are visible.}
    \label{fig:od_figure_more}
\end{figure*}
The bounds \eqref{eq : main result NESS} and \eqref{restdep:resderivatives} are also lower bounds on the EPR for overdamped Langevin dynamics. While they hold for coupled overdamped Langevin dynamics in arbitrary dimensions, we discuss the simplest case, a one-dimensional NESS, for which the dynamics in dimensionless form is given by
\begin{align}
  \dot{x} &= \mathcal{F}(x) + \eta(t). \label{odeq:Langevin}
\end{align}
Here, the force $\mathcal{F}(x)$ contains terms derived from a potential $V(x)$ and a non-conservative force $f$, and $\eta(t)$ is Gaussian white noise with $\langle\eta(t)\eta(t')\rangle = 2\delta(t-t')$. The steady-state EPR is $\sigma = \langle \mathcal{F}(x)\dot{x}\rangle$ \cite{shir23,seif25}.

For a first, simple illustration of the bound \eqref{restdep:resderivatives}, we consider a particle on a ring as sketched in Fig. \ref{fig:od_figure_more}. Here, the ring is completely observable. The dynamics of the particle is governed by the asymmetric potential
\begin{align}
  V(x) = -V_0\left[\sin(k_xx)-\frac{\sin(2k_xx)}{2}+\frac{\sin(3k_xx)}{3}\right] \label{odeq:potential}
\end{align}
with $k_x = 2\pi$ and a non-conservative force $f$. An observer, however, may have access to the two observables
\begin{align}
  A(x) &= 1.5\sin(k_xx) &\qq{and}&& B(x) &= 2.0\cos(k_xx) \label{odeq:observables}
\end{align}
yielding the entropy estimator \eqref{restdep:resderivatives}. Its quality factor \eqref{eq : quality factor tdep NESS} is shown for varying force $f$ and potential amplitude $V_0$ in Fig. \ref{fig:od_figure_more} (b).

Since the overdamped particle is in a NESS, the theoretical evaluation of all terms contributing to the estimator \eqref{restdep:resderivatives} is facilitated by the fact that we can express them as steady-state averages. We therefore move from the Langevin equation \eqref{odeq:Langevin} to its corresponding Fokker-Planck equation
\begin{align}
  \partial_t p(x,t) &= L_xp(x,t) = -\partial_x\left[\mathcal{F}(\re{x}) - \partial_x\right]p(x,t)
\end{align}
with mean local velocity $\re{\nu(x,t)}\equiv \mathcal{F}(x) - \partial_x\ln p(x,t)$. Using these equations and the fact that the system is in a NESS allows us to calculate its EPR as an average based on the force $\mathcal{F}$,
\begin{align}
  \sigma &= \re{\int_0^1} \mathcal{F}(x)\re{\nu(x,t)}p(x,t)\dd{x} = \langle \mathcal{F}^2 + \mathcal{F}'\rangle, \label{eq:odEPR}
\end{align}
where we drop the $x$ dependence and use a prime to indicate $\partial_x$. Moreover, for derivatives of correlation functions at vanishing lag, using $\partial_tp(x,t)$ leads to
\begin{align}
  \lim_{t\to 0}\partial_t\langle B(x_t)A(x_0)\rangle &= \langle (L_x^\dagger B)A\rangle = \langle (B'\mathcal{F} + B'')A\rangle, \label{eq:derivHelper}
\end{align}
where $L_x^\dagger$ is the adjoint of $L_x$.

Rewriting the time derivative of correlation functions as in Eq. \eqref{eq:derivHelper} grants access to where and why the quality factor reaches a value of $1$ in our first example in Fig. \ref{fig:od_figure_more} (b). In the special case of a flat ring, $V_0 = 0$, the estimator \eqref{restdep:resderivatives} equals the entropy production \eqref{eq:odEPR} if the observables $A(x)$ and $B(x)$ satisfy the conditions $A'\sim B$, $B'\sim A$ and $\langle A''B\rangle = 0 = \langle B''A\rangle$. These conditions are naturally met for observables that consist of one Fourier mode like $\sin(k_x x)$, and satisfy $A'\sim B$. More generally, this equivalence is approximately true for large enough forces $f$ for which the potential effectively vanishes.

Since full access of the ring allows for inferring the steady-state distribution and the force, see, e.g., Refs. \cite{fris20,das26}, we now slightly modify our first example to contain hidden regions, which prevents the reconstruction of the full potential. Such a coarse-graining depicted in Fig. \ref{fig:od_figure_more} (a) yields the quality factor of the bound \eqref{restdep:resderivatives} and its dependence on potential height $V_0$ and nonconservative force $f$ as shown in Fig. \ref{fig:od_figure_more} (c). Since in this example the three visible intervals make up half the ring, we infer roughly half the entropy production or less in this case.

Based on our knowledge of the one-dimensional topology in this simple example, we could also use waiting-time distributions or splitting probabilities based on sequences of observations to infer the full mean entropy production rate \cite{hart21a,meyb24}. However, the strength of the bound \eqref{restdep:resderivatives} is that we do not require measurements capturing sufficiently rare waiting times. Here, sufficient statistics for the short-lag limit of correlation functions and their derivatives is enough.

\section{Concluding perspectives}
In this work, we have derived two lower bounds on the EPR based on correlation functions of state observables applicable in a NESS or for time-dependent processes. To the best of our knowledge, these bounds are the first to use arbitrary coarse-grained state observables. Remarkably, the NESS bounds are valid for arbitrary correlation lags. Moreover, we only require a vanishing correlation lag for the EPR estimator that applies for arbitrary time-dependent dynamics. Since these bounds can be determined from observable correlation functions, they are directly operational, which gives them strong potential for broad use for experiments on small driven systems.

This work opens several promising research avenues. \re{As illustrated in Fig.~\ref{fig : figure diamond NESS}(b), the quality of the estimator decreases far from equilibrium. This loss originates from the inequality $(a-b)^2/(a+b)\le(a-b)\ln(a/b)/2$ used in the derivation, which is saturated only close to equilibrium. Qualitatively, this loss does not depend on the choice of observables and is shared by other estimators based on the second moments of currents, such as the thermodynamic uncertainty relation. A detailed analysis of these effects and of their interplay with coarse-graining is left for future work.} While we have discussed optimizing the estimator in a NESS with respect to constant shifts of observables, exploring more general transformations like linear or non-linear mappings could lead to even tighter bounds. Additionally, applying the EPR estimators to experimental data and assessing their robustness under limited statistics would be valuable. In experimental settings, coarse-graining may even be imperfect with noisy identification of meso-states from time series \cite{vdm25,bao25}. Investigating this additional layer of coarse-graining would therefore be interesting as well. In particular, since understanding how this faulty coarse-graining affects the quality of the estimators may allow us to compensate for these effects thus further increasing the potential of the estimators in experimental settings. Finally, we have required the vanishing-lag limit to derive the bound for time-dependent processes. Hence, it will be worthwhile to explore whether this bound can be extended to finite correlation lag, which could become relevant to experimental situations with limited time resolution.

\bibliography{references.bib}
\appendix

\end{document}